\documentclass[prl,superscriptaddress, notitlepage]{revtex4-1}

\usepackage[utf8]{inputenc} 
\usepackage[T1]{fontenc}    
\usepackage{hyperref}       
\usepackage{url}            
\usepackage{booktabs}       
\usepackage{amsfonts}       
\usepackage{nicefrac}       
\usepackage{microtype}      
\usepackage{lipsum}		
\usepackage{graphicx}
\usepackage{natbib}
\usepackage{lmodern} 
\usepackage{microtype}
\usepackage{natbib}
\usepackage{xcolor}

\begin{document}

\title{Retrieving optical parameters of emerging van der Waals flakes}

\author{Mitradeep Sarkar}
\author{Michael T. Enders}
\affiliation{ICFO-Institut de Ciencies Fotoniques, The Barcelona Institute of Science and Technology, 08860 Castelldefels (Barcelona), Spain}
\author{Mehrdad Shokooh‐Saremi}
\affiliation{Department of Electrical Engineering, Ferdowsi University of Mashhad,
Mashhad 91779-48944, Iran}
\affiliation{ICFO-Institut de Ciencies Fotoniques, The Barcelona Institute of Science and Technology, 08860 Castelldefels (Barcelona), Spain}
\author{Kenji Watanabe}
\affiliation{Research Center for Electronic and Optical Materials, National Institute for Materials Science, 1-1 Namiki, Tsukuba 305-0044, Japan}
\author{Takashi Taniguchi}
\affiliation{Research Center for Materials Nanoarchitectonics, National Institute for Materials Science,  1-1 Namiki, Tsukuba 305-0044, Japan}
\author{Hanan Herzig Sheinfux}
\author{Frank H.L. Koppens}
\author{Georgia Theano Papadakis*}
\affiliation{ICFO-Institut de Ciencies Fotoniques, The Barcelona Institute of Science and Technology, 08860 Castelldefels (Barcelona), Spain}

\begin{abstract}

High-quality low-dimensional layered and van der Waals materials are typically exfoliated, with sample cross sectional areas on the order of tens to hundreds of microns. The small size of flakes makes the experimental characterization of their dielectric properties unsuitable with conventional spectroscopic ellipsometry, due to beam-sample size mismatch and non-uniformities of the crystal axes. Previously, the experimental measurement of the dielectrirc permittivity of such microcrystals was carried out with near-field tip-based scanning probes. These measurements are sensitive to external conditions like vibrations and temperature, and require non-deterministic numerical fitting to some a priori known model. We present an alternative method to extract the in-plane dielectric permittivity of van der Waals microcrystals, based on identifying reflectance minima in spectroscopic measurements. Our method does not require complex fitting algorithms nor near field tip-based measurements and accommodates for small-area samples. We demonstrate the robustness of our method using hexagonal boron nitride and $\alpha$-MoO3, and recover their dielectric permittivities that are close to literature values.\\\\
\textbf{Email:}  georgia.papadakis@icfo.eu\\
\end{abstract}

\maketitle

\section{Introduction} 
\par{The spectral range of mid-long infrared (MLIR) light, pertaining to wavelengths $4\mu m-25\mu m$ is rapidly advancing as it is host to a broad range of applications in renewable energy \cite{Datas2019,Papadakis2021}, radiative cooling \cite{Fan2022}, thermal photon harvesting \cite{Xiao2022}, molecular sensing \cite{Bareza2022,Wang2023} IR spectroscopy \cite{John-Herpin2022}, thermal camouflage \cite{zhu2020lsa} and also for probing hot interstellar matter in our galaxy \cite{Draine2021,Packham2008}. In the search of MLIR relevant materials, recently discovered low-dimensional layered and van der Waals materials have nearly monopolized the scientific research  \cite{Zhang2020,Ermolaev2021}. In the MLIR, the coupling of lattice vibrations (phonons) with light yields phonon polaritonic modes \cite{Basov2016}, which are highly relevant for thermal photonics and spectrum engineering \cite{Caldwell2015}. As exemplary materials with MLIR phonon resonances, we mention classes of metal oxides (like $\alpha$-Mo$O_3$) \cite{Alvarez-Perez2020} as well as dielectrics, such as hexagonal Boron Nitride (hBN) \cite{Giles2018}, which are being explored for their extraordinary MLIR photonic properties.}
\\

\par{These van der Waals materials are dispersion-less at frequencies far from the phonon excitation bands. Near the phonon resonance frequency, these materials exhibit a dielectric permittivity that has a Lorentzian lineshape and changes sign rapidly with frequency. The permittivity near these resonances can be modeled with a standard Lorentz oscillator as:} 

 \begin{equation}
	\label{eq:hBNperm}
	\epsilon=\epsilon_{inf}\left[1+\frac{\omega_{LO}^2-\omega_{TO}^2}{\omega_{TO}^2-i\gamma\omega-\omega^2}\right]
\end{equation}       
\par{where $\epsilon_{inf}$, $\omega_{TO}$, $\omega_{LO}$ and $\gamma$ are the four parameters that characterize the material's macroscopic dielectric response. Within the so-called Reststrahlen band (between frequencies $\omega_{TO}$ and $\omega_{LO}$), these materials display anomalous dispersion \cite{Foteinopoulou2019}, hosting various interesting physical properties such as negative refraction \cite{sternbach2023s,hu2023s}, negative reflection \cite{alvarez-perez2022sa}, and others.}
\\

\par{Despite the rapidly growing interest in the properties of low-dimensional composites, standard approaches for the experimental characterization of their optical properties cannot be easily applied \cite{Yoo2022}. Although experimentally retrieving the frequency dispersion of the dielectric permittivity of a material required for \textit{any} photonic operation, obtaining the real and imaginary part of $\epsilon(\omega)$ for low-dimensional van der Waals materials is currently a very challenging task. Conventionally, the dielectric response of a materials is measured with Spectroscopic Ellipsometry (SE) \cite{Toudert2014}. Alternatively, one can use Fourier Transform Infrared Spectroscopy (FTIR), and, with certain assumptions on the thickness of a film and intrinsic material properties, via numerical fitting, the dielectric response can be obtained \cite{Alvarez-Perez2020}. Nonetheless, both SE and FTIR are optical techniques based on far-field optics, requiring hundreds of wavelengths wide (in the in-plane direction) sample areas for their accurate application. At MLIR frequencies, this calls for samples with cross sections on the order of millimeters owing to the long wavelengths of MLIR light. On the other hand, the cross-sectional sample area of van der Waals flakes that are exfoliated is typically on the order of tens to hundreds of micrometers \cite{Castellanos-Gomez2014}, and thus too small to be measured with FTIR or SE.}
\\

\par{Although different crystal fabrications techniques, like spin coating, chemical vapor deposition (rarely more than a few atomic layers thick)\cite{Ma2022}, van der Waal epitaxy \cite{Zeng2020} are reported as alternative approaches to exfoliaton, none can assure high crystal qualities over large surfaces till date. Furthermore, these materials are typically predicted to be highly anisotropic \cite{He2022}. Due to the random orientation of sub-crystals\cite{Niu2018}, high optical anisotropy is rarely observed in crystals with large surface areas, for which exfoliated flakes are widely considered as the ones with the highest crystal quality.} 
\\

\par{Previously, the experimental measurement of the dielectric permittivity of low-dimensional van der Waals flakes was carried out with near-field tip-based scanning probes \cite{Duan2022,Chaudhary2022}. These measurements are rather delicate and require a polarized narrowband laser \cite{DeOliveira2021} or broadband MLIR source \cite{Huth2011}. Furthermore, nano-scale imaging requires reconstruction of the hyperspectral data \cite{Li2015} which is not trivial.}
\\

\par{For an accurate retrieval of the real and imaginary part of the dielectric permittivity of a material, at least two observable quantities ought to be measured, for example the real and imaginary part of the reflection and/or transmission coefficients ($r$ or $t$) \cite{Smith2005} Alternatively, for example in SE, one measures the changes in the amplitude ($\psi$) and phase ($\Delta$) of the ratio of complex reflection coefficient for TM and TE polarized light ($r_P/r_S$) \cite{Yoo2022}. When the measurement of $r$ or $t$ is not possible, researchers rely on the real valued reflectance and transmittance ($R=|r|^2$ or $T=|t|^2$) to retrieve the complex permittivity \cite{Menzel2008,Kaplan2019}, either by elaborate numerical fitting algorithms \cite{Li2021,Lansford2020} or by fitting to numerically calculated spectra \cite{Niu2018,Lee2019,Jung2019}. We show here that this method based on one observable quantity ($R$ or $T$), on its own, is not robust in the case of low-dimensional crystals, due to experimental constraints related to flake sizes comparable to the wavelength of light, more so when phonon resonances occur in the spectral range of interest.}
\\

\par{In this work, we present a method to experimentally retrieve the in-plane complex dielectric permittivity of microscopic flakes of van der Waals materials from real valued reflectance measurements (via FTIR microspectrometry). The method relies on identifying the positions of the reflectance minima in the spectra caused by Fabry-Perrot (FP) resonances in the material. Such resonances provide direct access to deterministic evaluation of the real part of the refractive index of the material $Re\{n\}$ at the frequencies of the reflectance minima. Using multiple flakes of varying thicknesses, the real part of the refractive index can be determined over a wide spectral range. Our method, therefore, provided access to an FTIR microscope, provides access to the complex dielectric permittivity of low-dimensional exfoliated flakes using far-field optics.}
\\

\par{The characteristics of FP resonances have been recently used to determine the optical constants of lossless (or low-loss) materials \cite{Assali2016,Degli-Esposti2017}, or at frequency regimes far from the material resonances \cite{Lee2019}. Here, we demonstrate that, even in spectral regions of strong frequency dispersion, one can precisely evaluate the complex permittivity with minimal numerical fitting to evaluate $Re\{\epsilon\}$, utilizing FP resonances in thick van der Waals flakes. As a benchmark, we demonstrate the accuracy of our method by applying it to hBN (in-plane isotropic) and $\alpha$-Mo$O_3$ (in-plane anisotropic) within the spectral range of their respective Reststrahlen bands, in the frequency range of $600-8000cm^{-1}$ (limited by the frequency range of the FTIR used). We also demonstrate the in-plane anisotropy of $\alpha$-Mo$O_3$ and its hyperbolicity in the frequency range $600-813cm^{-1}$.}

\section{Experimental constraints and method} 

\par
\begin{figure*}[]
	\centering
	\includegraphics[width=1\linewidth]{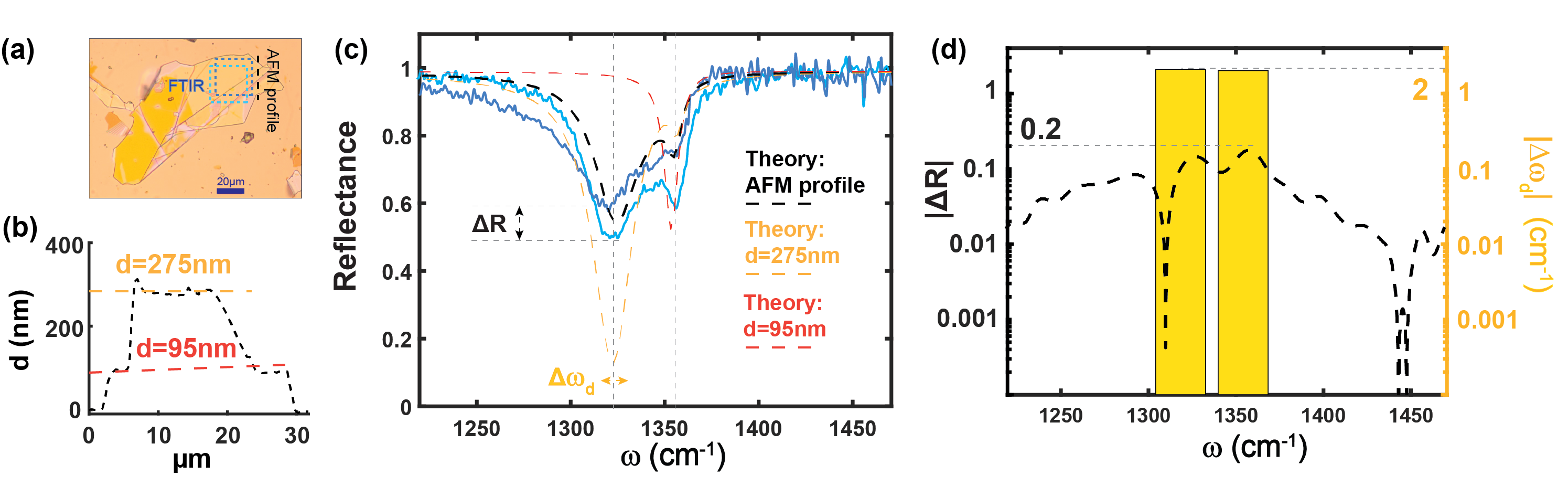}
	\caption{Microscope image of a hBN flake on gold. (b) Flake thickness (d) obtained by AFM along the black dashed line shown in a. Two distinct thickness are observed shown by dashed horizontal lines. (c) The reflectance spectrum obtained by FTIR microscope on two different locations of the flake (solid lines). The regions are marked in a using the same color code. The theoretical reflectance spectra are shown by dashed lines for d=275nm (yellow), d=95nm (red) and for the thickness profile shown in b (black). (d) The difference in reflectance ($\vert\Delta R\vert$) as a function of frequency for the two FTIR spectra shown in c (left axis). The maximum $\vert\Delta R\vert$ over the measured spectral range is 0.2. The frequency difference ($\vert\Delta\omega _d\vert$) of the reflectance minima for the same two FTIR spectra (right axis). $\vert\Delta\omega _d\vert$ for the two reflectance minima is 2$cm^{-1}$. }
	\label{fig:exptdisp} 
\end{figure*}

{A typical exfoliated flake of hBN on a gold substrate is shown in Fig. \ref{fig:exptdisp}a. The cross-sectional area of such flakes is roughly $\sim 100\mu m \times 100\mu m$. We also see that the thickness $(d)$ of the flakes are not uniform over the entire surface of the flake. The thickness used in this work were measured using Atomic Force Microscopy (AFM). A thickness profile on the hBN flake is shown in Fig. \ref{fig:exptdisp}b. Two distinct height distributions at $d=95nm$ and $d=275nm$ were observed for this particular flake. }
\\

\par{A standard reflectance spectrum, obtained via FTIR is shown in Fig.\ref{fig:exptdisp}c, taken on two different locations of the flake. By comparing the two measurements, there exists a high reflectance dispersion ($\vert\Delta R\vert$), owing to the non-uniformity of the flake, and such a discrepancy is very typical. More importantly, as the lateral dimensions of the exfoliated flakes are comparable to the wavelength of MLIR light, it results in significant diffraction and boundary effects.} 
\\
\par{The reflectance from a hBN film of given thickness d, on a gold substrate was analytically calculated and the spectra is superposed on Fig.\ref{fig:exptdisp}c for $d=95nm$ and $d=275nm$. The theoretical $R$ corresponding to a hBN layer height variation as shown in Fig.\ref{fig:exptdisp}b is also shown. The permittivity of the hBN retrieved by the method presented in this article was used for these calculations. We see that none of the analytical spectra matches the experimental curves. Hence, any attempt to retrieve the optical permittivity of hBN by the fitting the experimental reflectance spectra to numerical calculations is expected to result in large errors.}
\\
\par{On the other hand, the spectral position of the dips in reflectance ($\vert\Delta\omega_d\vert$) shows statistically much less variation with sample location, as also seen in Fig.\ref{fig:exptdisp}c. The two dips observed in the reflectance spectra correspond closely to hBN heights of $d=95nm$ and $d=275nm$. The maximum experimental discrepancy in $\vert\Delta R\vert$ was found to be $0.2$ whereas that for $\vert\Delta\omega_d\vert=2cm^{-1}$ for both the reflectance dips (Fig.\ref{fig:exptdisp}d). So retrieval of optical permittivity based on reflectance dip positions rather than fitting to the entire reflectance spectrum is more robust and precise.}
\\

\par
\begin{figure*}[]
	\centering
	\includegraphics[width=1\linewidth]{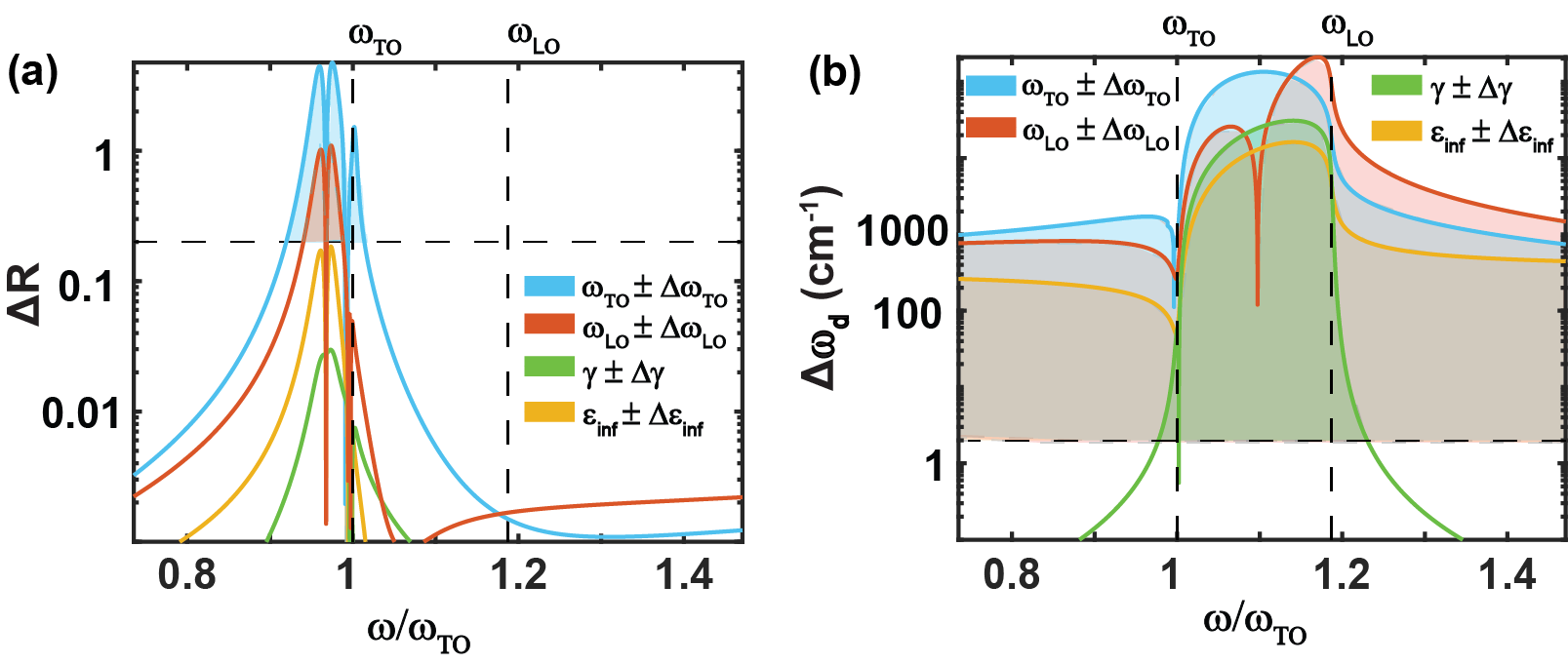}
	\caption{The reflectance from a material with Lorentz type permittivity and thickness (d) of 300nm on a semi-infinite gold substrate was calculated analytically. (a) $\vert\Delta R\vert$ as a function of frequency when any one of the $4$ parameters of the Lorentz model is changed by $\pm5\%$. The experimental precision ($\vert\Delta R\vert=0.2$) is shown by the horizontal dashed line. (b) $\vert\Delta\omega_d\vert$ when any one parameter is changed by $\pm5\%$. The experimental precision is considered to be $2cm^{-1}$. In this case all the 4 parameters of the Lorentz model can be determined with an error of less than $\pm5\%$.  }
	\label{fig:simuerror} 
\end{figure*}

{To emphasize this, analytical calculations were done to quantify the variation of $\vert\Delta R\vert$ and $\vert\Delta\omega_d\vert$ with respect to variations of the four material permittivity parameters ($\epsilon_{inf}$, $\omega_{TO}$, $\omega_{LO}$ and $\gamma$).  For hBN, the parameters were taken from Caldwell \textit{et al} \cite{Caldwell2014} (Table.\ref{tablehbn}). Each parameter was varied by $\pm5\%$ while the other 3 parameters were kept fixed. The reflectance spectra was analytically calculated for hBN ($d=300nm$) on gold substrate, and its variation as a function of $\omega$ is shown in Fig.\ref{fig:simuerror}a for the 4 parameters. $\vert\Delta R\vert$ is more than the experimental deviation of 0.2, only for $\omega_{TO}$ and $\omega_{LO}$. So, any method relying on $R$ measurements can determine $\omega_{TO}$ and $\omega_{LO}$ with an error of less than $\pm5\%$, while $\gamma$ and $\epsilon_{inf}$ cannot be accurately determined.}     
\\

\par{For any material on a highly reflecting substrate, the reflectance dips occur at frequencies where the FP resonance condition is satisfied. The frequencies ($\omega_d$) are related to the real part of the material refractive index ($Re\{n\}$) as \cite{Park1964}}
\begin{equation}
	\label{eq:refldip}
	\omega_d=\frac{1}{2d} \left[\frac{2m+1}{2Re\{n(\omega_d)\}}-\frac{1}{2\pi Re\{n(\omega_d)\}} tan^{-1} \left\{\frac{-2Re\{n(\omega_d)\} k_s)}{Re\{n(\omega_d)\}^2-n_s^2-k_s^2}\right\}\right]
\end{equation}
  
\par{where $d$ is the flake’s thickness, $m=0$ for first order resonances and $n_s$ and $k_s$ are the real and imaginary parts of the refractive index of the metal substrate respectively. $\vert\Delta\omega_d\vert$ with respect to one of the four parameters of the Lorentz model was analytically calculated using Eq.\ref{eq:refldip} and is shown in Fig.\ref{fig:simuerror}b. We see that $\vert\Delta\omega_d\vert$ is above the experimental deviation ($2cm^{-1}$) for all the four parameters of the Lorentz model.}
\\

\par{We identify $\omega_d$ from spectroscopic measurements. Then by using Eq.\ref{eq:refldip}, we evaluate the real part of the refractive index ($Re\{n\}$) of the material at frequency $\omega_d$. The reflectance spectra will have local minima for all frequencies at which Eq.\ref{eq:refldip} is satisfied. While $R$ is strongly dependent on both $Re\{\epsilon\}$ and $Im\{\epsilon\}$, thus requiring two observables for permittivity retrieval, $\omega_d$ is negligibly effected by $Im\{\epsilon\}$, hence only one observable suffice in extracting the complex optical permittivity. For clarity, we will group the reflectance dips and label them as $Res$ for the dips near the Reststrahlen bands of the material and as $FP$ for dips due to Fabry-Perot resonances at frequencies far from the material resonances.}

\section{Results and discussions}

\subsection{Retrieving complex permittivity of hBN }

\par
\begin{figure*}[]
	\centering
	\includegraphics[width=1\linewidth]{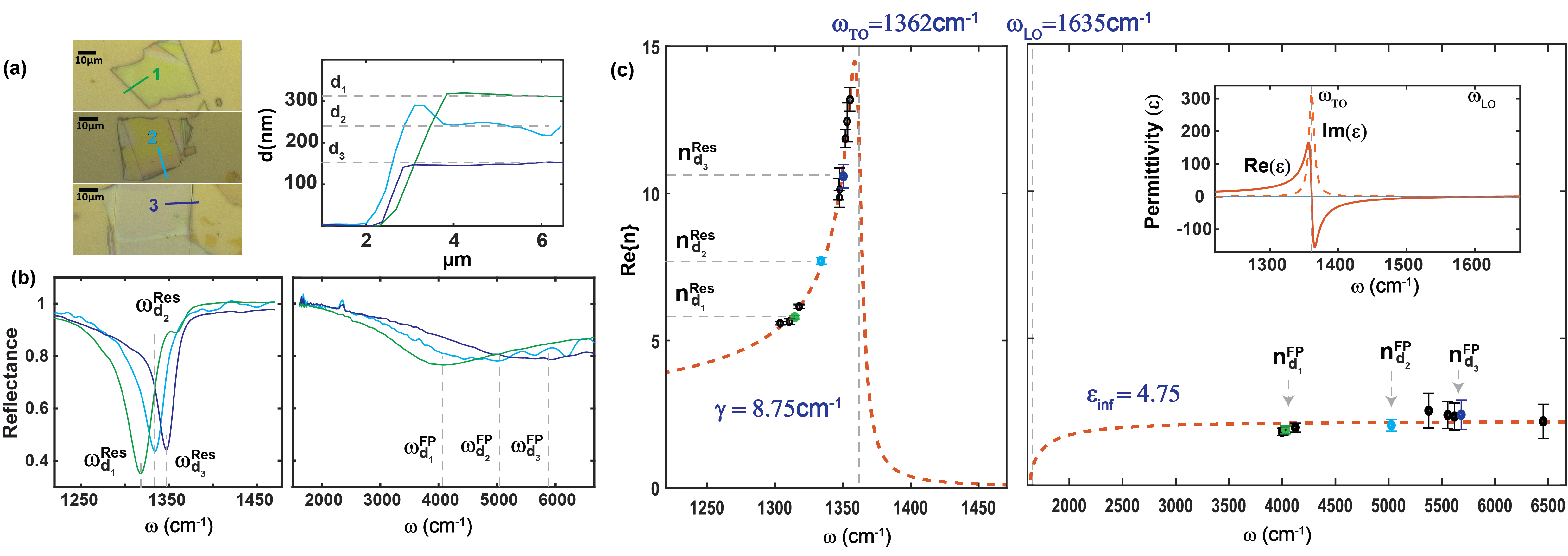}
	\caption{(a) Thickness profiles for 3 different hBN flakes on gold as measured by AFM. The microscope images of the 3 flakes are also shown and the positions where the profiles were measured are marked by lines with the same color code. The measured thicknesses (d) were 305nm, 225nm and 160nm respectively for flakes 1,2 and 3. (b) The reflectance spectra for the 3 flakes as measured by FTIR microscope is shown with the same color code as in a. The frequencies of the reflectance minima are marked as $\omega _{d_{1,2,3}}^{Res}$ for the dips close to the Reststrahlen band of the material and $\omega _{d_{1,2,3}}^{FP}$  for dips due to FP resonances at frequencies far from the material resonances. (c) Evaluation of $Re\{n\}$ of hBN at each $\omega _{d}$ measured for flakes of different thicknesses (black circles). The points corresponding to the 3 flakes shown in a and b are marked using the same color code. The Lorentz model was fitted to the $Re\{n\}$ to determine the 4 parameters and are shown in the figure. The evaluated real and imaginary parts of the permittivity of hBN as a function of frequency is shown as inset.}
	\label{fig:methodhbn} 
\end{figure*}

{The reflectance spectra for hbn flakes on gold substrate (11 flakes in total, 3 shown in Fig.\ref{fig:methodhbn}b) shows two first order ($m=1$ in Eq.\ref{eq:refldip}) dips, one close to the Reststrahlen band of hBN ($\omega _{d}^{Res}$) and the other far from the material resonances ($\omega _{d}^{FP}$). The corresponding $Re\{n\}$ were calculated for each reflectance dip and for each flake. These points for different flake thicknesses is shown in Fig.\ref{fig:methodhbn}c as black dots. The method of calculating the errors on the estimation of $Re\{n\}$ is shown in the Experimental section. The evaluated refractive index is almost dispersion-less far from the Reststrahlen band of hBN while the anomalous dispersion is observed near $\omega_{TO}=1361cm^{-1}$. }
\\

\par{Eq.\ref{eq:hBNperm} was fitted to the experimental $Re\{n\}$ to evaluate the four parameters and their respective standard errors (SE). They are $\omega _{TO}=1362\pm 1.6 cm^{-1}$,  $\omega _{LO}=1635\pm 29 cm^{-1}$, $\gamma=8.75 \pm 1.8 cm^{-1}$, $\epsilon _{inf}=4.75 \pm 0.35$. The root-mean-square error (RMSE) on the fit was $0.25$. The retrieved parameters are found to be very close to expected values for naturally occurring bulk hBN.}  
\\

\par{The reflectance dips close to the Reststrahlen band are narrow, hence $\omega _{d}^{Res}$ can be estimated accurately. This results in the low errors on the estimated values of $\omega_{TO}$ and $\omega_{LO}$. The reflectance dips far from the material resonance are broader and less pronounced. So, the estimated $\omega _{d}^{FP}$ are less precise. As these points are important to evaluate the high frequency permittivity,  $\epsilon_{inf}$, a higher standard error is observed in its value.} 
\\

\par{For accurate estimation of the damping factor, $\gamma$, data points are necessary in and close to the Reststrahlen band (Fig.\ref{fig:simuerror}b). However, the optical permittivity is negative in this band, and $Re\{n\}$ is close to zero. So, no reflectance dips occur in the Reststrahlen band. Also for frequencies lower than $\omega_{TO}$, but close to the Reststrahlen band, the reflectance dip is weak and is always not evident.  Hence, the estimated $\gamma$ is least precise, although the standard error is acceptable. The complex permittivity obtained by parameters retrieved by our method is shown in Fig.\ref{fig:methodhbn} as inset. The parameters for the complex permittivity of hBN is shown in Table.\ref{tablehbn} as obtained by our method and compared to values in the literature.}

\begin{table}[hbtp]
 \caption{Parameters for the in-plane complex permittivity of hBN}
 \label{tablehbn}
 \begin{center}
  \begin{tabular}{||c c c c c||}
    \hline
    Reference & $\omega _{TO} cm^{-1}$ & $\omega _{LO} cm^{-1}$ & $\epsilon _{inf}$ & $\gamma cm^{-1}$ \\
    \hline
    Zhao \textit{et al} \cite{Zhao2021} & $1370$ & $1610$ & $4.87$ & $5$ \\
    Caldwell \textit{et al} \cite{Caldwell2014}  & $1362.7$ & $1616.9$ & $4.98$ & $7.3$ \\
    Giles \textit{et al} \cite{Giles2018}  & $1360$ & $1614$ & $4.9$ & $7$ \\
    This work  & $1362$ & $1635$ & $4.75$ & $8.75$ \\
    \hline
  \end{tabular}
  \end{center}
\end{table}

\subsection{Retrieving complex permittivity of $\alpha$-Mo$O_3$ }
\par
\begin{figure*}[]
	\centering
	\includegraphics[width=1\linewidth]{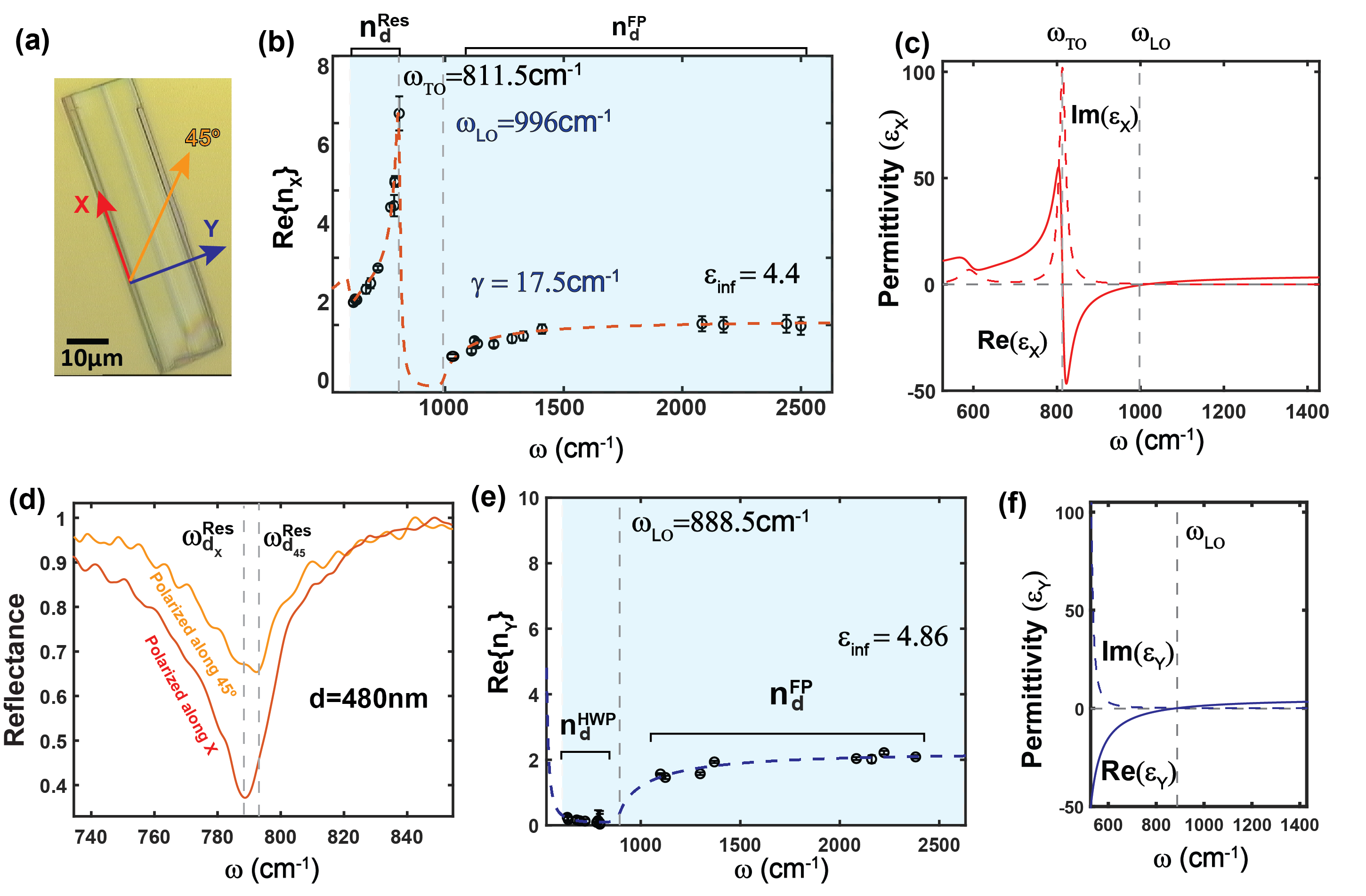}
	\caption{(a) Microscope image of a $MoO_3$ flake on gold. The two in-plane crystal axes (X,Y) are shown. (b) The determine evaluation of $Re\{n_X\}$ of $MoO_3$ at each $\omega _{d}$, measured for flakes of different thicknesses (black circles). The fit of the Lorentz model (red dashed) and the evaluated material parameters along the crystal axis $X$ are shown. (c) The evaluated permittivity of $MoO_3$ along $X$. (d) The reflectance spectra for a $MoO_3$ flake of d=480nm on gold when the polarizers are oriented along $X$ (red) and at 45$^{\circ}$ to $X$ (orange). The frequencies of the reflectance minima for the two curves are marked by vertical dashed lines. (e) The deterministic evaluation of $Re\{n_Y\}$ measured with the polarizers oriented along $Y$ for frequencies far from the material resonances ($n _{d}^{FP}$). For points in the Reststrahlen band of the material $n _{d}^{Eff}$ was calculated from the shift of $\omega _{d}^{Res}$ when the polarizers are oriented along X and along 45° to $X$ (as shown in d). The fit of the Lorentz model (blue dashed) and the evaluated material parameters along the crystal axis $Y$ are shown. (f) The permittivity of $MoO_3$ along $Y$.    }
	\label{fig:moo3} 
\end{figure*}

{The crystal structure of $\alpha$-Mo$O_3$ is orthorhombic which results in a strong in-plane anisotropy \cite{Zheng2023}. The permittivity can be decomposed into three principal values namely $\epsilon_X$, $\epsilon_Y$, and $\epsilon_Z$ along the three crystal directions, [100], [001] and [010] respectively. In this work, we retrieve the in-plane permittivity $\epsilon_X$ and $\epsilon_Y$. A typical $\alpha$-Mo$O_3$ flake, exfoliated from bulk crystal, on a gold substrate is shown in Fig.\ref{fig:moo3}a with the two in-plane directions marked. Due to the difference in the strength of the van der Waals forces along the three crystal directions, exfoliated $\alpha$-Mo$O_3$ flakes are generally rectangular in cross-section, with the long axis corresponding to the $X$ direction while the short axis corresponds to the $Y$ direction.}   

\par{Alvarez-Pérez \textit{et al} \cite{Alvarez-Perez2020} have recently reported the permittivity of $\alpha$-Mo$O_3$ from near and far-field correlative measurements. They used a Lorentz model for the permittivity, with three oscillators in $X$ and one oscillator in $Y$. The high frequency permittivities along $X$ was $\epsilon_{inf_X}=5.78$. The three Reststrahlen bands in $X$ are $\omega _{TO_1}=506.7 cm^{-1}$,  $\omega _{LO_1}=534.3 cm^{-1}$; $\omega _{TO_2}=821.4 cm^{-1}$,  $\omega _{LO_2}=963.0 cm^{-1}$ and $\omega _{TO_3}=998.7 cm^{-1}$,  $\omega _{LO_1}=999.2 cm^{-1}$. For the three resonances, $\gamma_1=49.1$, $\gamma_2=6.0$ and $\gamma_3=0.35$. The third oscillation is extremely narrow and the phonon resonance is extremely weak and this was not predicted in earlier theoretical \cite{Eda1991} or experimental studies \cite{Zheng2023,Li2018}. So we will ignore this resonance. The Lorentz model along $X$ is given as}  

 \begin{equation}
	\label{eq:moo3perm}
	\epsilon=\epsilon_{inf}\left[1+\frac{\omega_{LO_1}^2-\omega_{TO_1}^2}{\omega_{TO_1}^2-i\gamma_1\omega-\omega^2}\right]\left[1+\frac{\omega_{LO_2}^2-\omega_{TO_2}^2}{\omega_{TO_2}^2-i\gamma_2\omega-\omega^2}\right]
\end{equation}   

\par{Two linear polarizers, parallel to each other, one for the incident light and other for the reflected light was used. The axis of the polarizers were oriented along $X$ (long axis of the flake). The $Re\{n_X\}$ along $X$ was retrieved (using 13 flakes in total) and is shown in Fig.\ref{fig:moo3}b. The four parameters of the Lorentz model along $X$, are $\epsilon _{inf}=4.40 \pm 0.26$; $\omega _{TO_1}=589.5\pm 36 cm^{-1}$,  $\omega _{LO_1}=606.9\pm 25 cm^{-1}$, $\gamma_1=51.5 \pm 22 cm^{-1}$; $\omega _{TO_2}=811.5\pm 2 cm^{-1}$,  $\omega _{LO_2}=996.7\pm 12 cm^{-1}$, $\gamma_2=17.5 \pm 1.6 cm^{-1}$. The RMSE on the fit was 0.16.}
\\

\par{The FTIR is limited to the spectral range of $600-8000cm^{-1}$ while the first oscillator of the Lorentz model has its phonon resonance between $589.5cm^{-1}$ and $606.9cm^{-1}$ and hence for retrieval of its parameters, extrapolation is necessary. This results in the high errors associated with the parameters of the first oscillator, mainly for the damping factor $\gamma_1$. For the second oscillator of the model along $X$, all the three parameters were precisely determined by our method. The parameters of the second oscillator are summarized in Table.\ref{tablemoo3X} along with values from the literature. The complex permittivity along $X$ thus retrieved is shown in Fig.\ref{fig:moo3}c. }
\\

\begin{table}[hbtp]
 \caption{Parameters for the complex permittivity of $\alpha$-Mo$O_3$ along $X$}
 \label{tablemoo3X}
 \begin{center}
  \begin{tabular}{||c c c c c||}
    \hline
    Reference & $\omega _{TO} cm^{-1}$ & $\omega _{LO} cm^{-1}$ & $\epsilon _{inf}$ & $\gamma cm^{-1}$ \\
    \hline
    Zheng \textit{et al} \cite{Zheng2023} & $820$ & $972$ & $4$ & $4$ \\
    Alvarez-Pérez \textit{et al} \cite{Alvarez-Perez2020}  & $821.4$ & $963$ & $5.78$ & $6$ \\
    Li \textit{et al} \cite{Li2018}  & $818$ & $974$ & $-$ & $-$ \\
    This work  & $811.5$ & $996.7$ & $4.40$ & $17.5$ \\
    \hline
  \end{tabular}
  \end{center}
\end{table}

\par{For the permittivity along $Y$ only one oscillator is expected \cite{Alvarez-Perez2020}. reflectance spectra was obtained with the polarizer axes oriented along the short direction of the flakes ($Y$). As $\omega _{TO}$ for $\epsilon_Y$ is beyond the spectral range of the FTIR (Table.\ref{tablemoo3Y}), we cannot obtain reflectance dips corresponding to $\omega _{d}^{Res}$. So only dips corresponding to $\omega _{d}^{FP}$ were identified and $Re\{n_Y\}$ was evaluated for those frequencies. This is shown in Fig.\ref{fig:moo3}e.} 
\\

\par{Anisotropic crystals like $\alpha$-Mo$O_3$, cause a rotation of the polarization axis of incident light, polarized along $45^{\circ}$ to the two orthogonal principal axes of the crystal. This manifests as a dip in the reflectance measured with the two polarizers parallel to each other. The frequencies where this polarization rotation occurs is given by the same expression as Eq.\ref{eq:refldip} with $Re\{n\}$ replaced by the birefringence ($Re\{\Delta  n_{45}\}$) of the crystal given as}  

 \begin{equation}
	\label{eq:n45}
 Re\{\Delta n_{45}\}=\vert Re\{n_{X}\}-Re\{n_{Y}\} \vert      
\end{equation} 

\par{This is the half wave-plate (HWP) condition for anisotropic loss-less  materials.}
\\

\par{The polarizers were oriented along $45^{\circ}$ to $X$ direction and $\omega _{d_{45}}^{Res}$ were identified from the dips in the measured reflectance spectra. $\omega _{d_{45}}^{Res}$ is shifted with respect to $\omega _{d_X}^{Res}$ when the polarizers are along $X$ as shown in Fig.\ref{fig:moo3}d. From Eq.\ref{eq:refldip} we deduced the birefringence $Re\{\Delta n_{45}(\omega _{d_{45}}^{Res})\}$. From predetermined values of $n_X$ and Eq.\ref{eq:n45}, $Re\{n_{Y}\}$ was calculated. These points are grouped as $n _{d}^{HWP}$ in Fig.\ref{fig:moo3}e.}
\\

\par{The Lorentz model given by Eq.\ref{eq:hBNperm} was used for the permittivity along $Y$. The four parameters extracted from the fit are $\omega _{TO}=516.3\pm 62 cm^{-1}$,  $\omega _{LO}=888.5\pm 26 cm^{-1}$, $\gamma=18.06 \pm 8 cm^{-1}$, $\epsilon _{inf}=4.86 \pm 0.19$. The root-mean-square error (RMSE) on the fit was $0.10$. $\omega _{TO}$ is beyond the spectral range of the FTIR used, hence the low precision in determining the parameters $\omega _{TO}$ and $\gamma$. The complex permittivity along $Y$ is shown in Fig.\ref{fig:moo3}f and the parameters are summarized in Table.\ref{tablemoo3Y}.}

\begin{table}[hbtp]
 \caption{Parameters for the complex permittivity of $\alpha$-Mo$O_3$ along $Y$}
 \label{tablemoo3Y}
 \begin{center}
  \begin{tabular}{||c c c c c||}
    \hline
    Reference & $\omega _{TO} cm^{-1}$ & $\omega _{LO} cm^{-1}$ & $\epsilon _{inf}$ & $\gamma cm^{-1}$ \\
    \hline
    Zheng \textit{et al} \cite{Zheng2023} & $545$ & $851$ & $5.2$ & $4$ \\
    Alvarez-Pérez \textit{et al} \cite{Alvarez-Perez2020}  & $544.6$ & $850.1$ & $6.07$ & $9.5$ \\
    Li \textit{et al} \cite{Li2018}  & $545$ & $851$ & $-$ & $-$ \\
    This work  & $-$ & $888.5$ & $4.86$ & $-$ \\
    \hline
  \end{tabular}
  \end{center}
\end{table}

\par{Albeit, the limitations in evaluating precisely the $\omega_{TO}$ and $\gamma$ for the permittivity along $Y$, our value of $\omega_{LO_Y}=888.5cm^{-1}$ predicts that the permittivity along $Y$ is negative below this frequency (up to $600cm^{-1}$, which is the lower range of the FTIR). On the other hand, $\omega_{TO_X}=813.3cm^{-1}$, so the permittivity along $X$ is positive below this frequency. Hence, we can confirm that $\alpha$-Mo$O_3$ is hyperbolic in the frequency range $600-813.3cm^{-1}$. }

\section{Conclusion}

\par{In conclusion, in this work we showed that, using far-field FTIR spectroscopy in a microscope, we can accurately retrieve the dielectric permittivity of small (with respect to the beam size) van der Waals exfoliated flakes. In particular, near the MLIR phonon resonances of such materials, we retrieve all four parameters of the Lorentz oscillator model (Eq.\ref{eq:hBNperm}) with high accuracy. We show that, in previous approaches based on numerical fitting of reflectance measurements \cite{Kaplan2019,Lee2019}, the phonon linewidth ($\gamma$) as well as the background permittivity ($\epsilon_{inf}$), cannot be precisely determined by fitting numerical calculations to measured reflectance spectra. We show theoretically and demonstrate experimentally that the weak dependence of spectral position of a FP resonance in a dielectric slab ($\omega_d$ in Eq.\ref{eq:refldip}) on $Im\{\epsilon\}$ allows retrieval of the complex permittivity of van der Waals materials using far-field reflectance measurements.}
\\

\par{We have evaluated the complex permittivity of hBN and that of $\alpha$-Mo$O_3$ along both in-plane crystal directions. The values presented in this article match closely the theoretical and experimental permittivity previously reported in the literature. The method can be applied to any lossy, frequency-dispersive material, and does not require complicated experimental setup or near-field optics, nor complex numerical fitting algorithms. This method opens up new approaches for applying far-field optical techniques in van der Waals small-area exfoliated flakes.}


\section{Supplementary information - Methods}
\subsection{Exfoliated samples and characterisation} Gold of thickness $100nm$ was evaporated on glass substrates to serve as the reflector. The permittivity of gold (on glass) was measured by the IR-VASE Mark II (J.A. Woollam) ellipsometer for the spectral range of $333-8000cm^{-1}$. The complex refractive index of the gold was extracted using the CompleteEASE software (J.A. Woolam).

Both hBN and $\alpha$-Mo$O_3$ were mechanically exfoliated and
transferred onto the gold substrate with polydimethylsiloxane (PDMS) based exfoliation and transfer (X0 retention, DGL type from Gelpak) at 90$^{\circ}$C. The thickness of the exfoliated flakes were measured by Park NX20 Atomic Force Microscope using a non-contact cantilever (AC160TS, Olympus). 

The reflectance spectra were obtained using Bruker Hyperion 2000 microscope coupled to a Bruker Tensor FTIR spectrometer. The Mercury-Cadmium-Telluride (MCT) detector used has a spectral range of $600-8000cm^{-1}$. The light was focused on the sample and reflected light collected by a  Cassegrain objective ($\times36$, angular spread: $15^{\circ}-30^{\circ}$, numerical aperture: 0.5). For the reflectance from $\alpha$-Mo$O_3$,  two linear polarizers (ZnSe), parallel to each other were placed in the FTIR microscope, one for the incident light and other for the reflected light.   

The reflectance spectra were acquired with a spectral resolution of $2cm^{-1}$. An internal aperture of the microscope was adjusted to fit the part of the crystal where the reflectance is to be measured. All the reflectance spectra from the flakes were normalized to a reference gold mirror. 

The acquired spectra were filtered using cubic smoothing spline interpolation. Then the frequencies of the local reflectance minima ($\omega_d$) were identified manually. The fitting to extract the Lorentz model parameters was done by the least square method, with the model equation being $Re\{n\}=Re\{\sqrt{\epsilon}\}$ where $\epsilon$ being Eq.\ref{eq:hBNperm} or Eq.\ref{eq:moo3perm} as applicable.

\subsection{Theoretical reflectance calculations} Owing to the off-normal Cassegrain objective, which is used for the collection of the reflected light, the etendue of the objective must be accounted for when calculating reflectance. The reflectance measured along two orthogonal planes ($X$ and $Y$) of the objective, for wavelength $\lambda$ and incidence angle $\theta$, when the incident light is polarized along $X$ can be written as 

\begin{equation}\label{eq:conditions}
		\begin{array}{c c l}\displaystyle
		
		r_X = \frac{\left(1-\frac{cos(\theta_T)}{n_G cos(\theta)}\right) + i\left( \frac{\aleph_Z}{n_X^2cos(\theta)} - \frac{n_X^2cos(\theta_T)}{n_G \aleph_Z} \right) tan(d k_0 \aleph_Z )  } 
         {\left(1+\frac{cos(\theta_T)}{n_G cos(\theta)}\right) - i\left( \frac{\aleph_Z}{n_X^2cos(\theta)} + \frac{n_X^2cos(\theta_T)}{n_G \aleph_Z} \right) tan(d k_0 \aleph_Z ) }\\\\\displaystyle
		
r_Y = \frac{\left(\frac{cos(\theta)}{n_G cos(\theta_T)}-1\right) + i\left( \frac{\aleph_X}{n_G cos(\theta_T)} - \frac{cos(\theta)}{\aleph_X} \right) tan(d k_0 \aleph_X )  } 
         {\left(\frac{cos(\theta)}{n_G cos(\theta_T)}+1\right) - i\left( \frac{\aleph_X}{n_G cos(\theta_T)} + \frac{cos(\theta)}{\aleph_X} \right) tan(d k_0 \aleph_X )  }
  
			\end{array}
			\end{equation}

where the refractive indices are $n_X$ and $n_Z/\alpha_{Z}$ along the principal crystal directions $X$ and $Z$ respectively, $k_0=2\pi/\lambda$, $n_G$ being the complex refractive index of gold, $\aleph_Z=\sqrt{n_X^2-\alpha_Z^2sin^2\theta}$ and $\aleph_X=\sqrt{n_X^2-sin^2\theta}$. The total reflectance is then $R=(r_X r_X^*+r_Y r_Y^*)/2$. For the calculations $\theta$ was considered to be $22.5^{\circ}$, which is the central angle of the objective used. This was used to calculate the theoretical reflectance spectra of Fig.\ref{fig:exptdisp}c and for the calculations shown in Fig.\ref{fig:simuerror}. For the permittivity along $Z$, values from Caldwell \textit{et al}\cite{Caldwell2014} and Alvarez-Pérez \textit{et al}\cite{Alvarez-Perez2020} was used for hBN and $\alpha$-Mo$O_3$ respectively. 

We must note that with a metallic reflector, the reflectance is weakly dependent on the incident angle, so with $\theta=0$, the reflectance can be approximated accurately. This was used (Eq.\ref{eq:refldip}) to estimate $Re\{n\}$ from the reflectance dip positions.

\subsection{Estimating errors} For each flake, the error on the measured thickness ($\delta d$) and on $\omega_d$ ($\delta \omega_d$) were evaluated. The contribution of the two on the estimation of $Re\{n\}$ can then be calculated from Eq.\ref{eq:refldip} neglecting the second term as $\delta n_d=\delta d/(4d^2\omega_d)$ and $\delta n_\omega=\delta \omega_d/(4d\omega_d^2)$ respectively. Then the precision on the evaluation of $Re\{n\}$ was then calculated as $\sqrt{{\delta n_d}^2+{\delta n_\omega}^2}$. 

To evaluate the parameters of the permittivity, we performed a linear least square fitting. The $95\%$ confidence intervals ($CI$) for each of the parameters were calculated from the inverse R factor (of the QR decomposition of the Jacobian), the degrees of freedom for error, and the root mean squared error. The Standard Error is then $SE=CI/t$, where $t$ is the inverse cumulative distribution function and is $\approx 2.08$ for $20$ data points in the case of hBN, $\approx 2.07$ for $25$ data points for $\alpha$-Mo$O_3$ along $X$ and $\approx 2.11$ for $18$ data points for $\alpha$-Mo$O_3$ along $Y$.

\medskip
\textbf{Acknowledgements} \par 
This work is dedicated to the memory of John S.  Papadakis. The authors declare
no competing financial interest. The authors would like to thank V. Pruneri and J. Martorell
for generously granting access to their laboratories and
equipments. G.T.P.  acknowledges funding from “la Caixa”
Foundation (ID 100010434), from the PID2021-125441OA-I00 project funded by
MCIN/AEI/10.13039/501100011033/FEDER, UE, and from the European Union’s Horizon
2020 research and innovation programme under the Marie Skłodowska-Curie Grant
Agreement No. 847648. The fellowship code is LCF/BQ/PI21/11830019. This work is
part of the R\&D project CEX2019-000910-S, funded by
MCIN/AEI/10.13039/501100011033/, from Fundació Cellex, Fundació Mir-Puig, and
from Generalitat de Catalunya through the CERCA program.
M.E. acknowledges: Ayuda PRE2020-094401 financiada por MCIN/AEI/ 10.13039/501100011033 y FSE ``El FSE invierte en tu futuro''.
K.W. and T.T. acknowledge support from the JSPS KAKENHI (Grant Numbers 20H00354, 21H05233 and 23H02052) and World Premier International Research Center Initiative (WPI), MEXT, Japan.

\medskip

%
\bibliographystyle{MSP}
\bibliography{bib_extraction}


\end{document}